\documentstyle[aps,amssymb,twocolumn,epsfig]{revtex}
\def\be{\begin{equation}}
\def\ee{\end{equation}}
\def\bea{\begin{eqnarray}}
\def\eea{\end{eqnarray}}

\def\br{\begin{thebibliography}{abc}}
\def\er{\end{thebibliography}}

\def\inbar{\,\vrule height1.5ex width.4pt depth0pt}
\def\IB{\relax{\rm I\kern-.18em B}}
\def\IC{\relax\hbox{$\inbar\kern-.3em{\rm C}$}}
\def\II{\relax{\rm I\kern-.18em I}}
\def\IR{\relax{\rm I\kern-.18em R}}
\def\IZ{\relax\ifmmode\mathchoice
{\hbox{\cmss Z\kern-.4em Z}}{\hbox{\cmss Z\kern-.4em Z}}
{\lower.9pt\hbox{\cmsss Z\kern-.4em Z}}
{\lower1.2pt\hbox{\cmsss Z\kern-.4em Z}}\else{\cmss Z\kern-.4em Z}\fi}

\def\rt{\rightarrow}
\def\bighat{\widehat}
\newcommand{\sect}[1]{\setcounter{equation}{0}\section{#1}}

\def\sxn#1{\bigskip\medskip \sect{#1} \smallskip
                                                 }

\newpage

\begin{document}
\title{\bf A Novel Spin-Statistics Theorem in $(2+1)d$ Chern-Simons Gravity}

\author{A.P. Balachandran,\\
{\it Department of Physics, Syracuse University, Syracuse, NY
  13244-1130, USA} \and \\
        E. Batista,\\
{\it Universidade Federal de Santa Catarina, Centro de F\'{\i}sica e
  Matem\'atica, Dep MTM, CEP 88 010-970, Florian\'opolis, SC, Brazil}
\and \\
        I.P. Costa e Silva \and P. Teotonio-Sobrinho,\\
{\it Universidade de S\~ao Paulo, Instituto de F\'{\i}sica-DFMA, 
Caixa Postal 66318, 05315-970, S\~ao Paulo, SP, Brazil}}
%\date{Received: \quad September 15, 1998}
%\maketitle
%\address{$^{1}$Department of Physics, Syracuse University, Syracuse,
%NY 13244-1130, USA}
%\address{$^{2}$Universidade Federal de Santa Catarina, Centro de
%  F\'{\i}sica e Matem\'atica, Dep MTM\\
%Caixa Postal  , Florian\'opolis, SC, Brazil}
%\address{$^{3}$Universidade de Sao Paulo, Instituto de Fisica-DFMA,
%Caixa Postal 66318, 05315-970, Sao Paulo, SP, Brazil}
\address{\parbox{14cm}{\bigskip\rm\small
It has been known for some time that topological geons in quantum
gravity may lead to a complete
violation of the canonical spin-statistics relation : there may be no
connection between spin and statistics for a pair of geons.
We present an algebraic description of quantum gravity in a $(2+1)d$
manifold of the form $\Sigma \times \IR$,
based on the first order canonical formalism of general relativity. We identify
a certain algebra describing the system, and obtain its
irreducible representations. We then show
that although the usual 
spin-statistics theorem is not valid, statistics is completely
determined by spin for each of these irreducible
representations, provided one of the labels of these representations,
which we call {\it flux}, is superselected. We argue that this is
indeed the case. Hence, a new spin-statistics theorem can be formulated.
%\\ PACS numbers: 
%74.25.Nf,
%74.72.Bk,
%76.60.Es,
%74.20.Mn.
}}
\maketitle
\thispagestyle{myheadings}
%\markright{{\em SU-4240-704} \hspace{31mm}} 
%{\small submitted for publication in {\em Physical Review Letters} } \hfill \h%space{1mm}}
%\input amssymb.tex
%%%%%%%%%%%%%%%%%%%%%%%%%%%%%%%%%%%%%%%%%%%%%%%%%%%%%%%%%%%%%%%%%%%%%%%%%%%%%%%
\sxn{Introduction}\label{S1}

In general relativity, although the metric of spacetime is a dynamical entity
determined by Einstein's field equations, the underlying topology is
not {\em a priori} determined. On a closer inspection, however, one
actually finds that once one imposes that spacetime should possess
some physically reasonable geometrical conditions, the presence of
non-trivial topology is constrained. Simple examples are the
well-known constraints on the spacetime topology in
Robertson-Walker models. Also, in classical general relativity, when some
standard types of energy conditions are valid, non-trivial spatial
topology may lead to singularities in spacetime: Gannon's theorem
\cite{gannon} (see also \cite{lee}) implies that, in a spacetime satisfying
the weak energy condition, if one attempts to
develop Cauchy initial data on a spatial 3-manifold \footnote{More
  precisely, a partial Cauchy surface regular near infinity-see
  \cite{gannon} for the appropriate definitions.}  with a non-simply connected
topology, the corresponding Cauchy development will be geodesically
incomplete to the past or to the future. The so-called
active topological censorship theorem \cite{witt1} formulated more
recently states that in a
globally hyperbolic, asymptotically flat spacetime obeying an averaged
null energy condition (ANEC), every causal curve beginning and ending at the
boundary at infinity can be homotopically deformed to that
boundary. Therefore, an external observer near that boundary would not
be able to probe the non-simply connectedness of spacetime. This
result has been extended to more general contexts than the
asymptotically flat case, such as asymptotically anti-de Sitter spacetimes
(see \cite{witt2} and references therein).    

In spite of such results, there is still much room left for
investigation of the physical consequences of having a non-trivial spatial
topology, especially in quantum theory. On the one hand, even in the
classical case one can have non-trivial {\em compact} spatial
topologies, which evade the conditions of the above cited theorems
and also have
physical interest, and on the other hand, in quantum theory, the
energy conditions to prove these theorems are often violated: for
example Wald and
Yurtsever \cite{WY} show that ANEC is violated by the renormalized
stress tensor of free fields in generic curved spacetimes. Indeed, its
is the existence of this so-called quantum ``exotic'' matter that
permits the violation of the classical area theorem by evaporating black holes
\cite{candelas}, and the existence of ``traversable'' wormholes, in
spite of the above mentioned theorems (see, e.g., \cite{visser} for an
extensive account). Moreover, it is widely believed that
quantum gravity effects will alter the topology of spacetime at
Planck scales (``spacetime foam''). Indeed, some semiclassical
calculations indicate that a configuration with the presence of
wormholes is energetically favored over the euclidean one \cite{garattini}.

Topological geons, which are the subject of
this paper, are topological structures with some remarkable
properties. They were first
studied by Friedmann and Sorkin
\cite{FS}, as ``localized excitations of spatial topology'', or ``lumps''
of non-trivial topology in an otherwise Euclidean spatial background. The idea
was to view such entities as particles much in the same way as
solitons in a field theory. The presence of geons can give rise to
half-integer spin states and fermionic or even fractional statistics,
in pure (i.e., without matter) quantum gravity
\cite{FS,Sam}. It is common in the literature refer
to such solitonic states as geon states. We follow this usage here.  

Geons being soliton-like objects, we can talk about their spin and
statistics. In \cite{erice,aneziris}, it was shown that such states could
violate the usual spin-statistics theorem, in $(3 + 1)d$ and $(2 + 1)d$,
if the spatial topology is assumed not to change in time, or more
precisely if the topology of the spacetime $M$ is of the form
$M=\Sigma \times \IR$.      
On a spacetime of the form $M=\Sigma \times \IR$,
the topology of a spatial slice is well-captured by the geons on
$\Sigma$. For example, in the $(2 + 1)d$
context that we are interested in this paper, the topology of an
orientable, connected surface $\Sigma$ representing space, with at
most one asymptotic region, is
completely specified by the number of handles. Each handle corresponds
to a geon in this simple context. Accordingly, topology changes are
always associated with creation and annihilation of geons. It has been
suggested \cite{DS} that the standard spin-statistics relation
can be recovered if geons can be created and
annihilated, in other words, topology change may be required in
order to establish the full spin-statistics theorem for geons. In this paper we
seek instead a relation between spin and statistics assuming a {\em fixed}
spatial topology.

To appreciate the importance of having or not having a spin-statistics
connection for geons, one must recall that in ordinary 
quantum field theories in Minkowski spacetime, the particles which
arise when we second quantize, for example, have this connection
naturally. Now, in a hypothetical quantum theory of gravity, one could
think of geons as a ``particle'', representing the excitations
of the topology itself. It seems therefore natural to ask whether they
share this connection with normal particles. We find that in the
formalism we develop here a
different, weaker version of the spin-statistics connection arises,
instead of the normal one.

Before we describe our approach to this situation, we
examine more carefully what is
meant by spin and statistics. Let us assume that we have a
configuration space $Q$ describing a pair of identical geons. One
such configuration can be visualized as two handles on the
plane. Now, the quantization 
of two geons on the plane is not unique. One has to choose some
hermitian vector bundle
$B_k$ over $Q$ whose square-integrable sections (with a suitable
measure) serve to define the
domains of appropriate observables \cite{FS,erice,aneziris}, and are
the ``wave functions'' in the quantum theory. The
index $k$ labels inequivalent
quantizations. The space of these sections is the quantum
Hilbert space ${\cal H}_k$ of the two-geon system.  Physical
operations can be implemented as operators on
${\cal H}_k$. If we perform a $2\pi $-rotation of one of the geons,
described by an operator $C_{2\pi}$,
then its eigenstate will change by a  phase $e^{i2\pi S}$, where $S$
is the spin. Just like particles in $(2+1)d$, geons can
carry fractional spin, i.e, $S$ can be any real number \cite{aneziris,Sam}. 
Similarly, if we exchange the position of the
two geons, the wave function will change by the action of an operator ${\cal R}$
that we call the statistics operator. The standard spin-statistics
relation would tell us that the action of ${\cal R}$ on a two-geon
system should be equivalent to acting with the operator
$C_{2\pi}$ on one of the geons. Note
that there is no {\it a priori} reason for this relation to
hold since $C_{2\pi}$ and ${\cal R}$ correspond to two independent
diffeomorphisms of $\Sigma$. Now one can
ask if such a relation is true
for each quantization procedure parametrized by $k$. The results of
\cite{Sam,erice,aneziris,DS} shed some light on the problem. The
authors show that some quantizations
violate the spin-statistics theorem, but leave open the question of
which are the ones that do not. Furthermore, as emphasized in \cite{Sam}, 
the list of quantum theories derived in \cite{aneziris} is completely based
on kinematic considerations. In other words, only the diffeomorphism
constraint is imposed, whereas the Hamiltonian constraint, which gives
the dynamical features of gravity, is not considered at the quantum
level. Imposing
the latter would further restrict the states, and in this
sense some of the values of $k$ may not be dynamically allowed. 

In this letter we show that, at least for $(2+1)d$ gravity in the
first order formalism, there is a
generalization of the standard spin-statistics connection relating
${\cal R}$ and $C_{2\pi}$, even for a fixed spatial topology, i.e.,
for spacetime manifolds of the form $\Sigma \times \IR$. We shall 
consider $\Sigma$ to be a one-point compactified two-manifold, i.e.,
we compactify the spatial manifold with one asymptotic region by
adding a ``point at infinity''. In the
quantization scheme given in \cite{aneziris}, one considers the
mapping class group $M_\Sigma$ (the group of ``large''
{\em spatial} diffeomorphisms, not connected to the identity of
  $Diff(\Sigma)$) and finds
a vector bundle $B_k$ for each unitary irreducible representation of
$M_\Sigma$. Then, one  sees
no relation between ${\cal R}$ and $C_{2\pi }$ for a generic
$k$. The physical significance of this
procedure is as follows. Physical states in quantum gravity obey the
diffeomorphism constraint, meaning that they are invariant under
``small'' diffeomorphisms, i.e., the diffeomorphisms connected to the
identiy of $Diff(\Sigma)$, which are the ones generated by this
constraint. The diffeomorphism constraint means that ``small''diffeos
should be regarded as gauge, but leaves one free to
consider the states either as invariant under the ``large'' diffeos (those
not connected to the identity of $Diff(\Sigma)$), in which case the
``large''diffeomorphisms are also viewed as gauge, or just
``covariant'', i.e., transforming by an unitary representation of the
mapping class group. In this approach, ``large'' diffeos are
regarded as a symmetry of the theory. We adopt the latter view in
this work, the former being a special case of this view.

We will look at $M_ \Sigma$
as part of a larger algebra ${\cal A}$ of operators describing the
quantum theory of geons. It contains the group algebra of $M_\Sigma
$. Let us give an intuitive account of ${\cal A}$. We start by
considering the classical (reduced) configuration space $\tilde{Q}$ of
$(2 + 1)d$ gravity in the first order formalism which is based on the
$SO(2,1)$ gauge group. It is well-known that this is
the space of flat $SO(2,1)$ bundles over the space manifold
$\Sigma$. As we will discuss in more detail in the body of the paper,
this space admits a natural measure. The wave functions are then taken
to be square-integrable
functions with respect to this measure. We now describe the algebra
${\cal A}$ used for quantization. In building this algebra, we
consider only the minimum needed to investigate the spin-statistics
connection. First, we comment on
its general structure. Its first component
consists of the operators of ``position'' type on the space $\tilde Q$
and corresponds to the commutative algebra ${\cal F}(\tilde Q)$
of continuous functions of compact support $f:\tilde{Q} \rightarrow
{\Bbb C} $. Next we consider the operators corresponding to the
symmetries of the theory. The gauge group $SO(2,1)$ acting on $\tilde Q$
induces an action on functions. Again, instead
of $SO(2,1)$, we take its group algebra ${\cal G}$. Finally, we also
include the
algebra ${\cal U}$ of (suitable) remaining operators acting on 
${\cal F}(\tilde Q)$. In other words, ${\cal A}$ has the structure
\be
\label{thealgebra1}
{\cal A}=({\cal U}\otimes {\cal G})\ltimes {\cal F}(\tilde Q),
\ee
We then choose the algebra ${\cal U}$ to be the group algebra of
$M_{\Sigma}$. It contains all the 
operations necessary to investigate the spin-statistics connection.

Another important feature is that the
first order formalism naturally takes into account the dynamical constraints.
The possible quantizations are given by irreducible 
$\ast $-representations $\Pi_r$ of ${\cal A}$, where the index $r$
parameterizes
inequivalent quantizations. We show that there is a large class of
quantizations $\Pi_r$ such that statistics is totally
determined by spin according to the formula
\be \label{SSC}
\Pi_r({\cal R})=e^{i(2\pi S-\theta[r])}\II~,
\ee
on state vectors of spin $S$. Here the extra phase $\theta [r]$ is
completely 
fixed by the choice of the
representation $\Pi_r$.

The rest of the letter is organized as follows. In Section \ref{S2} we
briefly review
the first order formalism of general relativity and deduce the
classical configuration space and the group actions thereon. We then
proceed to the construction of the algebra. The geon algebra can be
viewed as an example of a {\em transformation group algebra}, first
studied by Glimm \cite{glimm}, and the representation theory of this algebra is
known. In Section \ref{S3} we analyze more closely the structure of
the algebra and classify the irreducible $\ast $-representations. We
then show how a class of
states in these representaions possess a spin-statistics connection,
namely those states
which are eigenstates of a certain charge operator. These states are
then argued to be the true physical states, due
to a superselection rule. We end the paper with some final remarks.

\sxn{The connection formalism}\label{S2}

In the first order formalism, one takes as fundamental variables a triad 
$e^{(3)a}=e^{(3)a}_\mu dx^\mu $, possibly degenerate and an $SO(2,1)$
connection one-form $A^{(3)a} = \frac{1}{2}\epsilon^{abc}\omega ^{(3)}_{\mu
  bc}dx^\mu$,
where $\omega ^{(3)a}_{bc}$ is the spin connection \footnote{In our
  notation, the superscipt (3) on the upper right denote fields on
  the three-dimensional spacetime $M$, of the form $\Sigma \times \IR$
and fields without superscipt
  correspond to their pullbacks to $\Sigma$.}. The
Einstein-Hilbert action takes the form
\be \label{EHA}
S=\int _M e^{(3)a}\wedge F^{(3)}_a + \mbox{boundary terms},
\ee
where $F^{(3)}_a = d_M A^{(3)}_a + \frac{1}{2}\epsilon _{abc}A^{(3)b}\wedge A^{(3)c}$ is the
usual curvature for the connection $A^{(3)}$. In our convention, Lorentz
spacetime indices are
represented by Greek letters, and spatial indices by Latin letters
$i,j = 1,2$. Internal $SO(2,1)$ indices are represented by Latin
letters $a,b = 0,1,2$. Boundary terms arise \cite{reggeteitel,brown} in
the cases in which the spatial manifold $\Sigma$ is non-compact, or compact
with bondary, and are of course zero for closed $\Sigma$.

Upon
variation of the action (\ref{EHA}) with respect to $A^{(3)}$ and $e^{(3)}$, we
find the equations of motion 
\bea
\label{eqmotion}
F^{(3)a} &=& 0; \nonumber \\
D_M e^{(3)a} &=& 0,
\eea
where $D_M$ denotes covariant differentiation with respect to the
connection $A^{(3)}$. Let us consider
the equations of motion (\ref{eqmotion}) in coordinates. Since $M$ is
taken to be of the form $\Sigma
\times \IR$, we can use a ``space + time'' splitting. 
We then obtain the following set of equations for the spatial
components:
\bea
\label{constraint}
F^{a}_{ij} &=& 0, \nonumber \\
D_{[i}e^{a}_{j]} &=& 0,
\eea
which are nothing but the pullback of the equations (\ref{eqmotion})
to $\Sigma$ by the natural inclusion $\Sigma \hookrightarrow \Sigma
\times \IR: x \mapsto (x,0)$. The covariant
differentiation is now with respect to the pullback $A$ of the
connection $A^{(3)}$. Note that eqs. (\ref{constraint}) do not involve
time derivatives of the basic fields: they are just constraints on
the fields $e^{a}$ and $A_a$ on $\Sigma$ at any given time, and
initial data are a set of basic fields on $\Sigma$ satisfying these
constraints. The remaining equations are the time evolution equations
for $e^{a}$ and $A_a$. Since we shall not make explicit use of the
latter, we omit them here.    

$A_{aj}$ and 
$\epsilon ^{ij}e^a_i$, $i=1,2$ are canonically conjugate variables
defined on $\Sigma $. The pairs $(e^{a}, A^{a})$ obeying the
constraints span the (reduced) phase space
${\cal P}$ of
the theory, which is just the cotangent bundle of the space of
$SO(2,1)$ connections on $\Sigma$. The canonical symplectic structure is
given by the Poisson brackets coming from (\ref{EHA}). The only
non-vanising ones are:
\be
\label{symplectic}
\{A_i^{a}(x), e^{b}_j(y) \} = \frac{1}{2} \delta _{ab} \epsilon_{ij}
\delta^{(2)}(x - y),
\ee
where $x,y \in \Sigma$. 

%The smeared (first class) constraints are:
%\bea
%\label{constraints2}
%C(\Lambda) &=& \int _{\Sigma} \Lambda _a F^{a}; \nonumber \\
%C'(\Omega) &=& \int _{\Sigma} \Omega _a De^{a}.
%\eea
%Here $\Lambda$ and $\Omega$ are smooth smearing functions with values in the
%dual of the Lie algebra of $SO(2,1)$. The functions $C(\Lambda)$
%and $C'(\Omega)$ are the infinitesimal generators of $SO(2,1)$ gauge
%transformations and infinitesimal spatial translations
%respectively. The latter are equivalent on shell to diffeomorphisms,
%as discussed by Witten \cite{witten2}. Under the Poisson bracket
%defined in (\ref{symplectic}), these functions form a closed Poisson algebra
%isomorphic to the Lie algebra of the $(2 + 1)d$ Poincar\'e group
%$ISO(2,1)$. Now, when $\Sigma$ is non-compact, in order to ensure
%differentiability of  $C(\Lambda)$ and $C'(\Omega)$ on the phase space, the smearing functions $\Lambda$ and
%$\Omega$ must vanish at infinity, which in turn implies that the theory must be
%invariant only by those diffeomorphisms and gauge transformations which are
%trivial at infinity.  
 
The quantum
theory in the ``position representation'' would be described by wave
functionals $\psi[A]$. The constraints can be easily
imposed before quantization, and one then quantizes only the
physical degrees of freedom. When $\Sigma$ is a closed (i.e., compact
and boundaryless) 2-surface, the
constraints imply \cite{Sam,witten2} that the physical configuration
space $Q$ is given by the moduli space of flat connections,i.e., the
set of equivalence classes of flat connections on $\Sigma$ under gauge
transformations. When $\Sigma$ is non-compact, however, one has to
specify how fields behave asymptotically. This choice gives rise to
boundary terms in (\ref{EHA}) \cite{reggeteitel,brown}, and the
physical configuration space is the space of those flat connections
which have the appropriate asymptotic behavior, modulo those gauge
transformations which preserve this behavior. 

The full analysis becomes considerably more complicated in the
non-compact case because of the asymptotic considerations involved. To
simplify matters we just perform a one-point-compactification of
$\Sigma$, by adding a point $p_{\infty}$, the ``point at infinity'',
since the boundary terms in (\ref{EHA}) will play no role
here. ``Rotations'' of geons will be considered to be about this
point, and we also fix a frame there. Thus, $\Sigma$ is topologically 
taken to be a closed suface with a marked point and a frame attached there.
 
%could just assume that $\Sigma$ is
%closed. However, this creates a technical problem: heuristically, since we are
%interested in talking about spin, we must introduce a diffeomorphism
%performing ``rotations''. These are ``large''diffeomorphisms. When one
%has a non-compact surface one may talk about ``rotations with respect to      
%infinity'', but in the closed case, there is no distiguished ``point
%of reference''. Mathematically, the corresponding
%element of the mapping class group of $\Sigma$ is trivial for closed
%$\Sigma$. We therefore take an intermediary scenario, by assuming that
%$\Sigma$ {\em is} closed, {\em but has a distinguished, marked point, which we
%denote by $p_{\infty}$, and rotations are considered to be about this
%point}. It can be chosen arbitrarily, but once fixed, it cannot be
%changed. We also fix a frame at this point. One can think of this point
%and the corresponding frame as mathematically representing an
%observer \footnote{One may alternatively think of this procedure as a
%  one-point {\em compactification} of a non-compact manifold $\Sigma$,
 % and $p_{\infty}$ as the ``point at infinity''. However, here we have
% neither a fixed metric nor specified fall-off rates on approaching
% $p_{\infty}$, in order to characterize this point as infinity, and
% hence one should be cautious in taking this view}. 

Again, just like in the usual closed case, the
configuration space is the space of all flat connections. However,
gauge transformations which are not trivial at infinity are {\em not}
a symmetry of the theory. Therefore, in our case, configurations which
differ by a gauge transformation which is not trivial at $p_{\infty}$
should not be viewed as equivalent. The
reduced configuration space in this case is theerefore the moduli
space space of flat connections modulo gauge transformations which are
trivial at $p_{\infty}$.   

Note also that we only need
regular flat initial data on $\Sigma$ to define the configuration
space $Q$, and to quantize. We make no assumption as to geodesic
completeness, and in particular, the formalism can accommodate geodesically
incomplete classical solutions. This is important, because in
classical general relativity, Gannon's
theorems \cite{gannon} imply
that singularities must
arise due to the multiple connectivity of $\Sigma$, at least when
$\Sigma$ is non-compact, under certain mild physical assumptions. Even
if the formation of singularities occurs in our case, this seems not
to interfere with the quantization procedure, at least
formally. On the other hand, precisely because of this independence,
it is not clear at this point what are the implications, if any, of
such singularities in the quantum theory.      

A connection $A$ on $\Sigma$ is determined by
its holonomies. For each closed curve $\gamma $ based at $p_{\infty}$ compute
the holonomy $W([\gamma])=P~e^{\int _\gamma A}$. This quantity is
invariant under gauge transformations that are identity at $p_{\infty}$. 
Since $A$ is flat, $W([\gamma])$ is
invariant under small deformations of $\gamma $ preserving $p_0$. In
other words, it depends only on the homotopy class $[\gamma ]$ of loop
$\gamma $. In fact, $W$
gives a homomorphism $\pi_1(\Sigma )\longrightarrow SO(2,1)$.

Let $\tilde Q$ be the set of all such maps. We recall that $W([\gamma])$
changes to $gW([\gamma])g^{-1},~~g\in SO(2,1)$, under gauge
transformations that are {\em not} identity (and equal $g$) at $p_{\infty}$. 
For closed surfaces with no marked point, one must make an
identification $W\sim gWg^{-1}$ to get the moduli space of flat
connections. In other words, $Q=\tilde Q/SO(2,1)$.
% We will use the phrase {\it
% gauge action} to refer to the action of $SO(2,1)$ on the holonomies.
%Instead of starting from a quantum theory on $Q$, 
%we can work with vector-valued wave functionals $\psi $ on the bigger space 
%$\tilde Q$, taking values in a vector space $V$, as it is done in covering space quantization \cite{balachandran}, 
%provided that $V$ carries a unitary representation $\rho$ of the 
%group action, i.e. , $\psi(gWg^{-1})=\rho(g)\psi(W)$. In this way
%quantities like probability densities $\psi^\dagger \psi$ are fully
%gauge invariant and can be regarded as observables. 
%We will call ``observable''
%any operator that commutes with the gauge action. 

In our case, $\Sigma$ is a two-dimensional surface
with a marked point $p_{\infty}$, which is chosen to be our base point. Gauge
transformations which are {\em not} trivial at $p_{\infty}$, taking
a value $g$ (say) at $p_{\infty}$, change $W$
to $gWg^{-1}$ as before, but, as explained, {\em these are no longer
  equivalent}. We call this action of $SO(2,1)$ by conjugation the
{\em gauge action}. It corresponds to a Lorentz transformation of our
chosen, fixed frame at $p_{\infty}$. The group
$Diff^{\infty}(\Sigma )$ of orientation-preserving {\em spatial}
diffeomorphisms (diffeos) which are trivial at $p_{\infty}$ (and leave
a frame there fixed) acts on
the holonomies $W$ by changing the curve
$\gamma $. Its subgroup $Diff_0^{\infty}(\Sigma)\subset
Diff^{\infty}(\Sigma )$, connected to the identity (the group of small
diffeos) cannot change the homotopy class of $\gamma $. 
Therefore the formulation is already invariant by small diffeos, and
the physical configuration space is $\tilde Q$.
Large diffeos, on the other hand, act nontrivially on the
holonomies. So, we can work with the quotient group $M_\Sigma
=Diff^{\infty}(\Sigma )/Diff_0^{\infty}(\Sigma)$, known as the
mapping class group. In particular, the elements $C_{2\pi }$ and  
${\cal R}$ are large diffeos\cite{FS,erice,aneziris}. 
%The whole formalism is obviously 
%invariant under $Diff_0(\Sigma )$, therefore 
For the sake of simplicity, we will denote the elements of
 $Diff^{\infty}(\Sigma)$
 and its  classes in $M_\Sigma $ by the same letters.
An important fact is that elements of  $M_\Sigma $ 
commute with the gauge action.

\begin{figure}[t]
\begin{center}
\begin{picture}(0,0)
%\put(0,0){0}
%\put(0,-50){-1}
%\put(0,-100){-2}
%\put(0,-150){-3}
%\put(50,0){1}
%\put(100,0){2}
%\put(150,0){3}
%\put(200,0){4}
%\put(-50,0){-1}
%\put(-100,0){-2}
%\put(-150,0){-3}
%\put(-200,0){-4}
\put(40,-10){$p_{\infty}$}
\put(40,0){$\ast$}
\put(20,54){$\gamma_1$}
\put(40,70){$\gamma_2$}
\put(96,44){$\gamma_3$}
\end{picture}
\epsfig{file=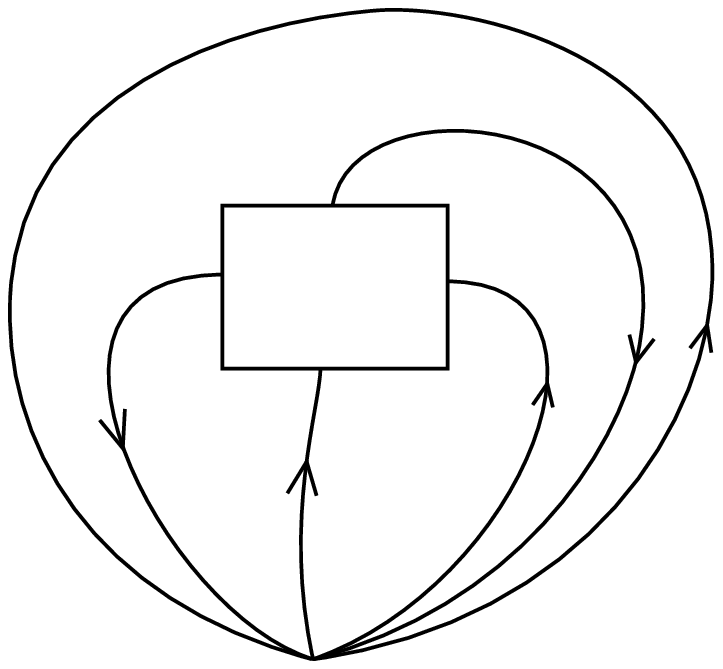,height=3cm}
\end{center}
{\small {\bf Fig. 1:} The figure shows $\Sigma$ for a single geon
  (opposite sides of the rectangle are to be identified) and loops
  $\gamma_i$ ($1 \leq i \leq
  3$). The homotopy classes $[\gamma_1]$ and $[\gamma_2]$ generate the
  fundamental group, while $[\gamma _3]$ is not independent of
  $[\gamma_1]$ and $[\gamma_2]$.}
\end{figure}

\sxn{The Geon Algebra}\label{S3} 

The algebra ${\cal A}$ used for
quantization has the structure
\be
\label{thealgebra}
{\cal A}=({\cal U}\otimes {\cal G})\ltimes {\cal F}(\tilde Q),
\ee
%If we take ${\cal U}$ to be generated by finite translations on the
%space $Q$ (translations on $\tilde Q$ that commute with the gauge
%action), then ${\cal A}$ would be the analogue of the Weyl algebra. Instead of
%doing so, we will take an effective theory where only some observables
%are included in ${\cal U}$. 
%For our purpose it is enough to take ${\cal U}$ to be
%the group algebra of the mapping class group, since it contains the
%operations necessary to investigate the spin-statistics connection. At
%the end we argue that the results derived from the effective theory
%holds for the complete one.
where ${\cal G}$ is the group algebra of $SO(2,1)$ and ${\cal F}(\tilde
Q)$ is the space of complex-valued, continuous functions with compact
support on $\tilde Q$. We choose the algebra ${\cal U}$ to be the
group algebra of $M_{\Sigma}$. ${\cal A}$ contains all the 
operations necessary to investigate the spin-statistics connection. 

Let us give an explicit presentation of ${\cal A}^{(1)}$, the algebra
${\cal A}$ for a single geon. We choose the
generators of $\pi_1 (\Sigma )$ to be the homotopy classes
of the loops  $\gamma _1$ and $\gamma _2$ of
Fig.1. Each flat connection provides us with  a pair of holonomies 
$(a,b)=(W(\gamma _1),W(\gamma _2))$. Since there are no relations among
the generators of $\pi_1(\Sigma )$, any pair of values $(a,b)$ can
occur. Therefore $\tilde Q$ is $SO(2,1)\times SO(2,1)$.

%The physical configurations do not span {\it a priori} all of $\tilde
%Q$, since
%we still have to impose boundary conditions. These must encode the
%well-known fact that in $(2 + 1)d$ the metric needs only be
%conical, rather than minkowski, at spatial infinity. It can be shown
%\cite{Sam} that imposing this condition amounts to taking the
%configuration space to be spanned by the pairs $(a,b)$ subject only to the
%condition $aba^{-1}b^{-1} \in SO(2) \subset SO(2,1)$. We will however
%proceed in another way. We will work throughout with the space $\tilde
%Q$ and will impose this condition on the representations of the geon algebra.
%Now, 

Instead of working with ${\cal F}(\tilde Q)$ directly, we work with
one of its representations. Note that the
Haar measure on $SO(2,1)$ induces a measure on $\tilde Q$. Using
this measure we may define an inner product on ${\cal F}(\tilde Q)$ in
the obvious way. The completion of ${\cal F}(\tilde Q)$ in this norm
is a Hilbert space ${\cal H}_0$, which is the space of
square-integrable functions (with this measure) on $\tilde Q$, carrying what we
call the defining representation of ${\cal F}(\tilde Q)$. A 
function $f\in {\cal F}(\tilde Q)$ acts on $\varphi \in {\cal H}_0$ as a 
multiplication operator:
\be
(f\varphi)(a,b)=f(a,b)\varphi(a,b)
\ee

With $g\in SO(2,1)$, let $\hat{\delta}_ g$ denote the generators of the group 
algebra $\cal G$. These $\hat{\delta} _g$'s  are gauge
transformations, and act by 
conjugating holonomies:
\be
(\hat{\delta} _g \varphi)(a,b)=\varphi(g^{-1}ag,g^{-1}bg)
\ee

The mapping class group of
$\Sigma $ has two generators $A$ and $B$, which correspond to Dehn
twists along the loops. Their effect on loops $\gamma _1$ and $\gamma_2$ is 
given by
%\begin{figure}
%\epsfbox{ab.eps}
%{\bf Fig. 2:}{\it \small The action of Mapping Class Group on loops.}
%\end{figure}
\be
\begin{array}{l}
(A\varphi)(a,b)=\varphi(a,ba^{-1}),\\
(B\varphi)(a,b)=\varphi (ab^{-1},b)
\end{array}
\ee 

The generators of ${\cal A}^{(1)}$ are functions
$f\in {\cal  F}(\tilde Q)$, diffeos $A,B$ of the mapping class group and gauge 
transformations $\delta _g$. 
%They satisfy the following relations
%\be
%\label{rel3}
%\begin{array}{ll}
%f_1f_2=f_2f_1, &\delta _{g_1}\delta _{g_2}=\delta _{g_1g_2},\\
%A\delta _g=\delta _gA, &B\delta _g=\delta _gB.
%\end{array}
%\ee
%Together with the relations
%\be
%\label{rel1}
%\begin{array}{l}
%(\delta_g f \delta_{g^{-1}} )(a,b) =f(g^{-1} ag,g^{-1} bg)\\
%(AfA^{-1})(a,b)=f(a,ba^{-1}) \\
%(BfB^{-1})(a,b)=f(ab^{-1} ,b) .
%\end{array}
%\ee

The mapping class group includes $C_{2\pi}$
\cite{FS,erice,aneziris,x1}. Its action on the
defining representation is
%\begin{figure}
%\epsfbox{c2pi.eps}
%{\bf Fig. 3:}{\it \small The operator $C_{2\pi}$}
%\end{figure}
\be
(C_{2\pi} \varphi )(a,b)=\varphi (cac^{-1} ,cbc^{-1})
\ee 
where $c:=aba^{-1}b^{-1}$. One can verify that $C_{2\pi} =(AB^{-1}A)^4$.  

These operators can be encoded in what is
called a transformation
group algebra \cite{glimm}. Let $G$ be a group with a left-invariant
measure acting on a space $X$. The
transformation group algebra is just the set of continuous functions 
${\cal F}(G\times X)$, with compact
support and with the product 
\be
(F_1F_2)(g,x)=\int _G~F_1(z,x)F_2(z^{-1}g,z^{-1}x)dz.
\ee
Here $x \rightarrow z^{-1} x$ is the group action on $X$, $z^{-1}g$
is the group product of $z^{-1}$ and $g$, and $dz$ is the left-invariant
measure on $G$. The irreducible representations of a transformation
group algebra have been worked out in \cite{glimm}. In our 
case, $X=\tilde Q$ and $G=SO(2,1)\times M_\Sigma $, where $G$ can be
made into a topological
group by giving $M_{\Sigma}$ the discrete topology. The measure on
$SO(2,1)$ is the Haar measure and the measure on $M_{\Sigma}$ is given
by 
\[
\sum _{m \in M_{\Sigma}} f(m)
\]
for any function $f$ on
$M_{\Sigma}$ with appropriate convergence properties. The measure on
$G$ is then the product measure. Finally,
${\cal A}^{(1)}={\cal F}(SO(2,1)\times M_\Sigma \times \tilde Q)$,
where we use the bijection 
\be
\IC(G) \otimes {\cal F}(X) \Longleftrightarrow {\cal F}(G \times X)
\ee
by interpreting $\delta _g \otimes f$ as the distribution
\bea
\delta _g \otimes f : \;\; (h,x) &\mapsto & \delta _g(h)f(x) \\
\nonumber 
&\equiv & \delta (g,h)f(x)
\eea
on $G \times X$, $\delta _g$ being the $\delta$-function supported at $g$. 

Let $Y=\tilde Q/G$ be the set of orbits of $G$ in $\tilde Q$, one such
orbit being ${\cal O}_{\omega}$. Let us choose one representative 
$(a_{\omega} ,b_{\omega}) \in \tilde Q$ for each orbit ${\cal
  O}_{\omega}$, and write ${\cal O}_{\omega} =
[(a_{\omega},b_{\omega})]$. We define
the stabilizer group $N_{\omega} \subset G$ as the set of elements
$(g,\lambda)$ of $G$ such that $(g,\lambda )\cdot (a_{\omega} ,b_{\omega} )=
(a_{\omega} ,b_{\omega} )$, where the $G$ action has been denoted by a
dot. Let $\alpha $ be a unitary irreducible representation of $N_{\omega}$ on
some Hilbert space $V_\alpha $. Now consider the space of
square-integrable functions 
$\phi :G\rt V_\alpha $ such that 
$\phi (hg,\xi \lambda )=\alpha (g^{-1}, \lambda^{-1})\phi(h,\xi)$
for all $(g,\lambda)\in N_{\omega}$ and $(h,\xi)\in G$. They are
called equivariant
functions. The set of these functions can be completed into a Hilbert
space $L^2(G,V_ \alpha)$ \cite{glimm}. The irreducible unitary
$\ast$-representations $\Pi_{(\omega ,\alpha)}$ of 
${\cal F}(G \times \tilde Q)$ can be realized on the Hilbert spaces 
${\cal H}_{(\omega ,\alpha)}=L^2(G,V_ \alpha)$ and, up to 
unitary equivalence, labeled by $r=(\omega ,\alpha )$. This label is a
quantum number characterizing a single geon. The
action of the operators $\hat F=\Pi_{r}(F)$, $F \in {\cal A}^{(1)}$ on
a vector $\phi ^r\in {\cal H}_r$ is given by
\begin{eqnarray}
\label{action}
(\hat F\phi^r )(h,\xi) &= \int _{SO(2,1)\times
  M_{\Sigma}}~F((h,\xi)\cdot (a_{\omega} ,b_{\omega} ), (g,\lambda))
  \times 
\nonumber \\
&\phi^r(g^{-1}h,\lambda^{-1} \xi )dz,
\end{eqnarray}
for any $h \in SO(2,1)$ and $\xi \in M_{\Sigma}$. We find, in
particular, that
\be
\begin{array}{l}
\left(\widehat{\delta _{h'}}\phi^r\right)(h,\xi)=
\phi^r(h'^{-1}h,\xi)\\
\left( \widehat{A}\phi ^r\right)(h,\xi)=\phi^r(h,A^{-1}\xi)\\
\left( \widehat{B}\phi ^r\right)(h,\xi)=\phi^r(h,B^{-1}\xi)\\
\left( \widehat{f}\phi ^r\right)(h,\xi)=
f\left(h\xi\tilde q_y\right)\phi^r(h,\xi).
\end{array}
\ee 

Now, let $\Sigma$ be an orientable surface of
genus two with a marked point $p_{\infty}$. It supports a system of
two geons. Their algebra ${\cal A}^{(2)}$ can be presented in the defining
representation space ${\cal H}_0\otimes {\cal H}_0$ of ${\cal
  A}^{(1)}\otimes {\cal A}^{(1)}$. It is generated by
elements of ${\cal A}^{(1)}\otimes {\cal A}^{(1)}$ plus the elements of the
mapping class group that mix up the geons, with the proviso that we
retain only ``diagonal''elements of the
form $\delta _g \otimes \delta _g$ from the gauge
transformations. There are only two independent generators of  $M_\Sigma$
involving both geons. One of them, the
diffeo ${\cal R}$ that exchanges the position of the
geons, has already been discussed in connection with the
spin-statistics relation. The other one is the so-called handle slide
$H$. Unlike the exchange ${\cal R}$, the handle slide $H$ has no analogue
for particles. Its existence comes from the fact that a geon is an
extended object. As the name indicates, it corresponds to the
operation of sliding an end of one of the handles through the other handle.
%Its action on  ${\cal H}_0\otimes {\cal H}_0$ is
%\begin{figure}
%\epsfbox{soumhandleslide.eps}
%{\bf Fig. 4:}{\it \small The handle slide.}
%\end{figure}
%\begin{eqnarray}
%\label{RH}
%&H&(\varphi (a_1 ,b_1 )\otimes \varphi (a_2 ,b_2 )=\nonumber\\
%&=& \varphi (a_1 (a_2 b_2 a_2^{-1}),(a_2 b_2 a_2^{-1})^{-1} 
%b_1(a_2 b_2 a_2^{-1}))\otimes \nonumber\\
%&\otimes & \varphi ((a_2 b_2 a_2^{-1})^{-1} 
%b_1(a_2 b_2 a_2^{-1}) a_2, b_2 )
%\end{eqnarray}

Our description of a pair of geons should be given by an algebra 
${\cal A}^{(2)}$ which also includes $H$. {\it But since $H$ does not
  enter directly in the
spin-statistics relation, we will not include it in ${\cal A}^{(2)}$}.

Although 
${\cal  A}^{(1)}$ is not a Hopf algebra, there is an element 
$R\in {\cal A}^{(1)}\otimes {\cal A}^{(1)}$ that plays the role of an
$R$-matrix. In other words, we can write ${\cal R}=\sigma R$ where 
$\sigma :{\cal H}_0\otimes {\cal H}_0\rt {\cal H}_0\otimes {\cal H}_0$
is the flip automorphism 
$\sigma \left(f_1\otimes f_2\right)=f_2\otimes f_1$. The $R$-matrix 
turns out to be
\be
\label{rmatrix}
R=\int \int da~db~P_{(a,b)}\otimes \delta ^{-1}_{aba^{-1}b^{-1}},
\ee 
where $P_{(a,b)}(\tilde q,h,\xi)=
\delta\left(\tilde q,(a,b)\right)\delta(h,e)\delta(\xi,e)$, the
$\delta$'s being $\delta$-functions. The existence of the $R$-matrix
is essential to establish the connection between spin and
statistics. It relates a diffeo performed on a pair of objects with
operators acting on each object individually. 

Each geon carries a representation ${\cal H}_r$ labeled by quantum numbers
$r = (\omega,\alpha)$. However, we only need to consider eigenstates of 
$\hat C_{2\pi}:=\Pi^{r}(C_{2\pi})$ with spin $S$. Let 
$\{\phi ^{r,S}_i\}$ be a basis for the eigenspace of spin $S$ in
${\cal H}_r$ for some fixed $r$. Two geons are said to be identical if
they carry the same quantum numbers $r$ and $S$. We
consider identical geons, fix an element
$(a_{\omega} ,b_{\omega})$ in the corresponding class $\omega$ and denote 
the net flux $a_{\omega} b_{\omega}
a_{\omega}^{-1}b_{\omega}^{-1}$ by $c_{\omega}$. Consider the
characteristic function $P_c$ which at $(a,b)$ is 1 if $aba^{-1}b^{-1} = c$
and zero otherwise. It is clear that a generic vector $\phi ^{r,S}_i$
is not an eigenstate of $\hat P_c$. A simple computation shows that
$\phi ^{r,S}_i$ is an eigenstate of $\hat P_c$ if and only if it has
support only on points $(h,\xi)$ such that $hc_ \omega h^{-1} =
c_{\omega}$.  

The quantum state for two identical 
geons is a linear combination of vectors of the
form $\phi ^{r,S}_i\otimes \phi ^{r,S}_j$. It is enough to show
the spin-statistics connection (\ref{SSC}) for such decomposable
vectors. We must
act with the operator $\bighat{{\cal R}} = (\Pi_r \otimes \Pi_r )({\cal
  R})$ on these vectors. By using eq. (\ref{action}), we easily see
that 
\be
\label{display1}
\bighat{P}_{(a,b)} \phi^{r,S}_i (h,\xi ) = 
\delta((a,b), (h,\xi)\cdot (a_{\omega} ,b_{\omega} )) \phi^{r,S}_i(h,\xi )
\ee
for every $(h,\xi ) \in SO(2,1) \times M_{\Sigma}$. Also, 
\be
\label{display2}
\bighat {\delta}_{c^{-1}} \phi^{r,S}_j (h,\xi ) =
\phi^{r,S}_j(ch,\xi ),
\ee
 where we have put $c=aba^{-1}b^{-1}$. Using
(\ref{rmatrix}) and the flip automorphism we conclude that
\begin{eqnarray}
\label{first}
&\bighat{{\cal R}}\phi ^{r,S}_i(h_1 ,\xi_1 )\otimes 
\phi ^{r,S}_j(h_2 ,\xi_2 ) =\nonumber\\ 
&={\bighat \delta}_{h_2 c^{-1}_{\omega} h^{-1}_2} \phi^{r,S}_j (h_1 ,\xi_1) \otimes 
\phi ^{r,S}_i(h_2 ,\xi_2 ).
\end{eqnarray}

{\em At this point we make the assumption that $\phi^{r,S}_{i,j}$ are
eigenstates of the net flux $\hat P_c$}, explaining its physical
meaning later. So we can set
$h_2c_{\omega}h^{-1}_2=c_{\omega}$. But we have
\begin{eqnarray}
\label{c2pi}
\bighat{\delta}_{c^{-1}_{\omega}}\phi^{r,S}_j (h_1 ,\xi_1) &=& e^{i2\pi S}
\bighat{\delta}_{c^{-1}_{\omega}} {\bighat C_{2\pi}}^{-1}
\phi^{r,S}_j (h_1,\xi_1)=\nonumber\\
&=& e^{i2\pi S}\phi^{r,S}_j (c_{\omega} h_1 ,C_{2\pi} \xi). 
\end{eqnarray}

Note that $\phi^{r,S}_j (c_{\omega} h_1 ,C_{2\pi} \xi) = \phi^{r,S}_j
(h_1 c_{\omega} , \xi C_{2\pi})$ because of the above assumption, and
because  $c_\omega$ commutes with $h_1$ and $C_{2\pi}$ commutes with every
element of $M_\Sigma$. On the other hand, $(c_\omega ,C_{2\pi})\in
N_\omega$ and
hence we can use the equivariance property of $\phi^{r,S}_j$ to
rewrite the  r.h.s. of the last equality in (\ref{c2pi}) as 
\[
\phi^{r,S}_j (c_{\omega} h_1 ,C_{2\pi} \xi) =\alpha (c_{\omega}^{-1} ,C_{2\pi}^{-1})
\phi^{r,S}_j (h_1 , \xi) .
\]
Now, every $\delta _g$ commuting with $a_\omega$ and $b_\omega$
commutes also with $c_\omega$, while $C_{2\pi}$ is in the center of
$M_{\Sigma}$. Therefore, $(c_\omega ,C_{2\pi})$ is in the center
of $N_\omega$, and by Schur's lemma we conclude that 
$\bighat{\delta_{c^{-1}_\omega}} {\bighat C_{2\pi}}^{-1}$ is equal to a
phase, say $e^{-i \theta (r)}$. Eq. (\ref{SSC}) then follows:
\be
\label{sscagain}
\bighat{{\cal R}}\phi ^{r,S}_i\otimes \phi ^{r,S}_j = e^{i[2\pi
  S-\theta(r)]}\phi ^{r,S}_j\otimes \phi ^{r,S}_i.
\ee  
%\sxn{Discussion}\label{S4}

We were able to establish a connection between spin and statistics for
all eigenstates of the net flux $\hat P_c$. In other words, {\em a
  spin-statistics exists for states with a definite net flux}. Now why
are these states special? The answer is that other
vectors in the representation space of $r$ are not physically allowed as a
consequence of a superselection rule, which we will discuss below. As a
consequence, only vectors which are in the eigenspace, say
${\cal H}_c$, of $\hat P_c$ are to be viewed
as pure quantum states. Linear
combinations of vectors in different ${\cal H}_c$'s are not pure, much
in the same way as one cannot have pure states
of different charges in QED, for example.

This superselection is actually very natural. First, note that the
net flux of a geon
commutes with all elements of the algebra except the gauge
transformations at $p_{\infty}$. Now, the gauge action cannot
be viewed as having a local effect from the standpoint of the geons, their
effect being limited to performing a transformation on the frame at
infinity. The other operators,
like those corresponding to the
mapping class group operators are ``local'', in the sense that they
correspond to operations on the geons themselves, i.e., operations
which can be taken to leave the region outside some ball surrounding the
geons invariant (no other, stronger
notion of locality is possible here, since we have no fixed background
metric). This is mathematically reflected in the fact that all
elements of the geon algebra other than the gauge
transformations (which are ``local''in the above sense) themselves
commute with the gauge action.

Therefore, given some eigenspace ${\cal H}_c$ of a net flux operator
$\hat P_c$, all operators other than gauge transformations preserve ${\cal
  H}_c$. Only the gauge transformation, say corresponding to an element $g \in
SO(2,1)$, takes vectors in ${\cal H}_c$ into vectors in ${\cal
  H}_{gcg^{-1}}$. That is, gauge transformations do change the net
flux, but this change does not correspond to a physical, local operation in
the theory; rather it is merely a relabeling of the fluxes. Once one fixes
the frame, and considers only local operations, one concludes that
the net flux can be regarded
as a charge which commutes with all the local operators, and hence is
superselected.  

\sxn{Final Remarks}\label{S7}

In this paper, we have shown a relation between the actions of the
diffeomorphisms $\hat C _{2\pi}$ and ${\cal R}$ on a class of geon
staes in $(2 + 1)d$ quantum gravity. An algebra describing the system
was identified and its representations were explained in detail. 

Our discussion can be viewed a generalization of previous work \cite{x1,x2},
where a spin-statistics relation was derived for geonic states arising
in a Yang-Mills theory coupled to a Higgs field in the Higgs phase,
where the symmetry is spontaneously broken down to a finite gauge
group $H$. In \cite{x2} we showed the existence of a class of ``localized''
states in quantum gravity arising indirectly from the Yang-Mills
theory which did obey the spin-statistics relation derived
here. However, those states form a very restricted class. The present
paper greatly expands the scope of the original version to a much larger class of geonic states in quantum gravity.   

In our version of the spin-statistics relation, there appears an
extra phase $\theta _r$ for each representation, and a natural
question is what is its meaning. It turns out to be a somewhat involved
problem, which we are presently tackling \cite{construction}.

\vspace{.5cm}
\noindent
{\large \bf Acknowledgments}

\noindent
We thank J.C.A. Barata for useful discussions, and S. Carlip,
J. Louko, D. Marolf, J. Samuel. R. Sorkin, and D. Witt for important
comments regarding the existenc of geons. The work of
A.P. Balachandran was supported by the Department of
Energy, U.S.A., under contract number DE-FG02-85ERR40231. The work of
E. Batista, I.P. Costa e Silva and
P. Teotonio-Sobrinho was supported by FAPESP, CAPES and CNPq
respectively.

\end{document}